# Metasurface-based Terahertz Three-dimensional Holography Enabled by Physics-Informed Neural Network

*Jingzhu Shao, Ping Tang, Borui Xu, Xiangyu Zhao, Yudong Tian, Yuqing Liu, and Chongzhao Wu**

J. Shao, P. Tang, B. Xu, X. Zhao, Y. Tian, Y. Liu, C. Wu
Center for Biophotonics
Institute of Medical Robotics
School of Biomedical Engineering
Shanghai Jiao Tong University,
Shanghai 200240, China
*Email: czwu@sjtu.edu.cna

**Abstract:**

Artificial intelligence has revolutionized optical device design, overcoming the efficiency bottlenecks of traditional methods. For holographic metasurfaces, conventional iterative algorithms suffer from time-consuming iterations and convergence stagnation, especially as the complexity of 3D target fields increases. While recent deep learning-based algorithms have improved the trade-off between speed and image quality, most existing models remain constrained by predefined physical scenarios (e.g., fixed distances), limiting their adaptability in dynamic, practical applications. To address these challenges, we propose a physics-informed neural network (PINN) based on local polynomial fitting and multi-plane wave propagation (LM-PINN) for the rapid design of terahertz 3D holographic metasurfaces. By leveraging a self-supervised training strategy, LM-PINN eliminates the need for labeled datasets, enabling direct end-to-end mapping from target holographic patterns to the metasurface structures. Both simulated and experimental results demonstrate that LM-PINN-designed metasurfaces offer higher imaging quality than traditional iterative algorithms. Crucially, by incorporating a distance encoding process, a single trained LM-PINN generalizes effectively across diverse physical configurations, including varying diffraction distances and distinct 2D or 3D targets, eliminating the necessity for retraining. Furthermore, the inference process of LM-PINN typically takes less than 1 second, providing a multifold speed advantage over traditional algorithms. Consequently, this strategy offers a robust and universal framework that paves the way for high-quality, real-time, and large-scale 3D holographic technologies.

## 1. Introduction

Computer-generated holography (CGH) enables the recording and reconstruction of light fields by numerically calculating holographic interference patterns (phase information) and encoding them into surface structures or spatial light modulators (SLMs).[1,2] Compared to conventional optical holography, CGH eliminates the reliance on physical interference systems and real objects, enabling a broader application. Currently, CGH plays a key role in various fields, including augmented reality (AR), virtual reality (VR), 3D display, and data storage.[3–6] In the terahertz (THz) wave band, although SLMs have been used to generate THz holograms,[7,8] current THz spatial modulation technology is still limited by the weak interaction between traditional optical materials and THz waves.[9] Commercially available products have not been widely adopted yet, which restricts the practical applications of THz SLM.[10]

Metasurfaces are artificially designed planar optical elements composed of periodic meta-atoms at subwavelength scales.[11–13] The elaborate shapes, materials, and arrangement of these meta-atoms endow them with unique capabilities to manipulate the amplitude, phase, polarization, and other properties of electromagnetic waves, enabling numerous novel functional devices such as multifocal lenses, compact spectrometers, and special beam generators.[14–16] Recently, metasurface-based holography has enabled the generation of diverse holograms ranging from 2D planes to 3D volumes,[17,18] exhibiting broad potential applications in communications, data storage, and display. Meanwhile, holographic metasurfaces operating in the THz band also demonstrate unique value in scenarios including subwavelength-resolution light manipulation, 3D projection, and encrypted display.[9,19–21]

As the functionality of holographic metasurfaces becomes increasingly complex, the difficulty of the design process also rises, especially in the THz band, where the available optical materials and devices remain relatively underdeveloped.[22,23] Conventional numerical calculation methods bear substantial computational burdens while balancing calculation speed and imaging quality. As a representative traditional iterative algorithm, the Gerchberg-Saxton (GS) algorithm converges the phase distribution of the metasurface through multiple round-trip propagations of light fields between the metasurface and the target imaging plane.[24] Though being intuitive in concept, there are limitations associated with the GS algorithm: computational time surges sharply as the size of the metasurface increases, and it only adapts to phase-only metasurfaces. Neglecting amplitude often compromises imaging quality, whereas complex amplitude modulation, while superior for 3D and multi-color displays, significantly exacerbates design complexity.[25,26] Representative THz complex-amplitude metasurfaces are summarized in Note S1. Traditionally, the design of complex-amplitude metasurfaces relies on matching the target field distribution of each pixel to predefined meta-atom libraries that enable independent phase and amplitude control,[19,27] a process that nevertheless necessitates extensive trial-and-error searches and remains computationally intensive. Furthermore, the overall electromagnetic response of the metasurface may not be a simple superposition of individual meta-atom responses, but rather a collective effect, especially when inter-unit electromagnetic coupling exists.[28]

With the rapid development of artificial intelligence (AI), deep neural networks (DNNs) have injected new vitality into the inverse design of metasurfaces.[29,30] Leveraging DNNs' data-driven advantages, DNNs can establish an accurate mapping relationship between metasurface geometric structures and electromagnetic response (or target imaging patterns) through massive training data, thereby enabling end-to-end design of holographic metasurface.[27,28] However, such supervised DNNs are essentially regarded as "black-box" models, which lack physical interpretability. It is difficult to clearly link optical principles with the design of metasurface structures in such a framework, which is unfavorable for subsequent optimization and mechanism exploration. Moreover, the performance of DNNs is highly dependent on large-scale paired input-label datasets. The data collection, preprocessing, and model training require significant time investment. Especially in THz wave bands, high-performance sources and detectors are under development,[23,31–33] and obtaining substantial amounts of high-quality labeled experimental data is challenging. Existing studies have shown that integrating physical models into unsupervised or self-supervised DNNs can not only reduce reliance on paired input-label datasets and improve network training efficiency but also enhance the interpretability of the model and design accuracy with the help of prior physical knowledge,[18,34–39] opening up a new technical path for the efficient and precise design of holographic metasurfaces. In unsupervised architectures, the network parameters are usually randomly initialized, and the optimal metasurface design is obtained through continuous iteration.[18,34–37] The advantage of this method lies in its high flexibility: when the physical scenario changes (e.g., the distance of target holographic imaging is altered), the parameters in the physical model can be directly modified to adapt to the new situation. In contrast, networks with a self-supervised architecture can be trained using a dataset of target holographic patterns (without the need for labels), enabling end-to-end design.[38,39] However, they suffer from a typical "one-model-one-scenario" limitation, where their physical parameters are generally fixed. When the optical path changes, the model needs to be retrained to meet the new design requirements.

In this work, we propose a local polynomial fitting and multi-plane wave propagation based physics-informed neural network (LM-PINN) for the rapid inverse design of THz 3D multi-plane holographic metasurfaces. To enhance the model's generalizability across diverse physical scenarios, we incorporate distance encoding into the input module. This allows a single trained LM-PINN to facilitate the end-to-end design of metasurfaces for varying target projection distances. LM-PINN utilizes a differentiable local polynomial fitting process as a physical surrogate instead of a separate forward prediction network,[37,38] significantly reducing the parameter count and training cost while ensuring high inference speed and accuracy. The feasibility of LM-PINN was validated through single-plane and multi-plane holographic scenarios, where it outperformed the Gerchberg-Saxton (GS) algorithm in FDTD simulations (SSIM, NPCC, and PSNR as the metrics). Furthermore, metasurfaces designed via LM-PINN were fabricated using lithography and etching, with experimental verification conducted via a THz quantum cascade laser-based holographic system. The experimental results align closely with FDTD simulations, confirming the practical feasibility of the proposed method. Building upon the validated framework, the LM-PINN integrated with distance encoding, termed Dist-LM-PINN, exhibits remarkable versatility. We demonstrate that a single Dist-LM-PINN, although trained on 2D datasets, enables the rapid design of metasurfaces not only for various target patterns at different distances, but also for multi-focal metalenses and complex 3D holograms. Notably, the trained LM-PINN achieves an inference time of approximately 1 second for a 512 × 512 array on a standard six-core CPU, significantly faster than the GS algorithm. This work will facilitate high-quality, real-time, and large-scale 3D holographic display in the THz band and provide a valuable reference for the AI-empowered design of THz devices.

## 2. Result and discussion

### 2.1. Architecture of the framework

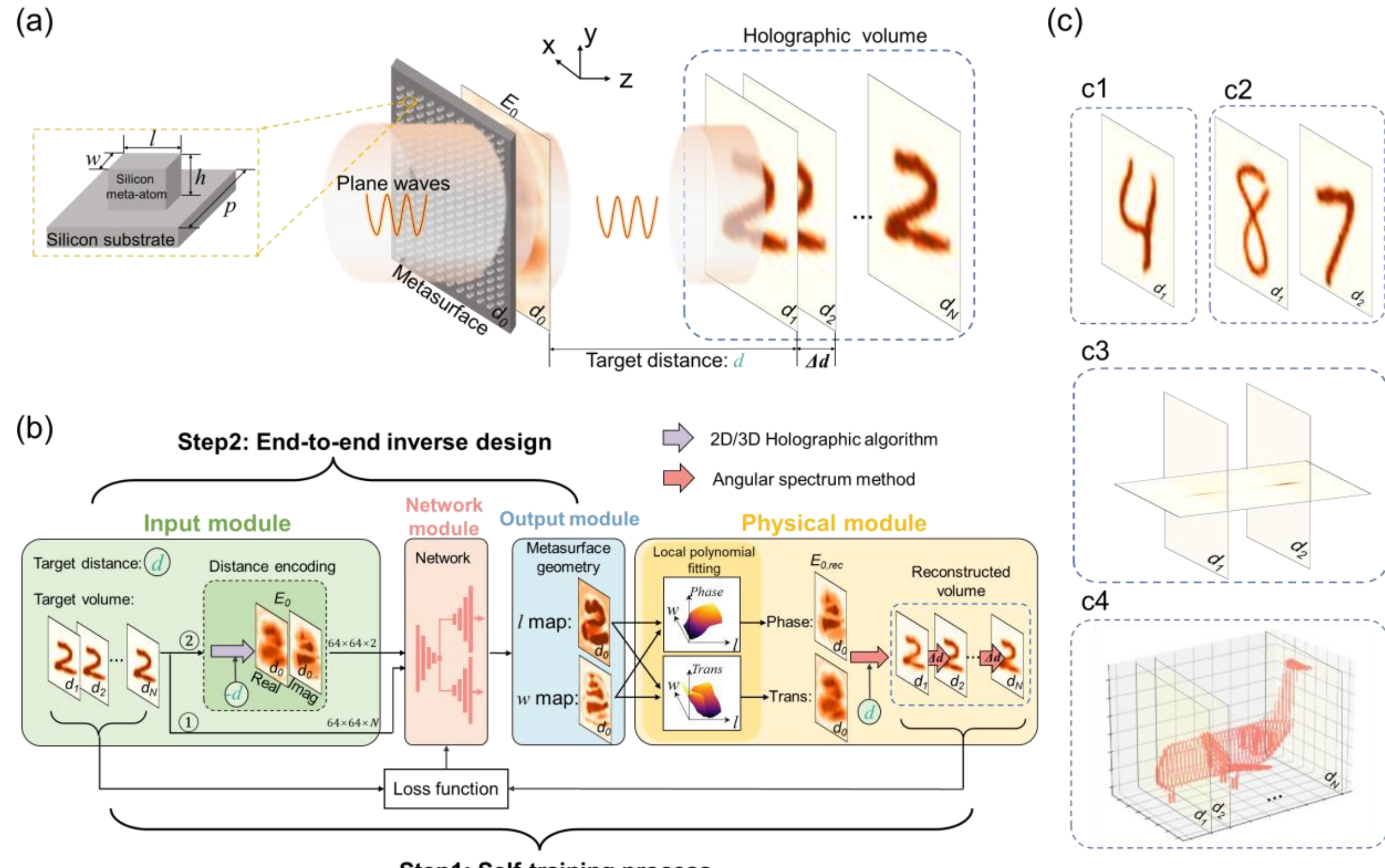

**Figure 1.** The framework for metasurface-based THz 3D holography. (a) Schematic of metasurface-based THz 3D multi-plane holography. The enlarged view shows the width ($w$), length ($l$), height ($h$), and period ($p$) of the silicon meta-atom. (b) The workflow of LM-PINN. (c) Different types of the holographic volume: c1 for single-plane target image, c2 for dual-plane with different target images, c3 for multi-foci, and c4 for 3D plane model.

**Figure 1**a illustrates the schematic of the all-dielectric metasurface generating 3D multi-plane holograms. A THz plane wave ($\lambda$ = 116 μm, $f$ = 2.58 THz) illuminates the metasurface, which consists of a 64 × 64 array of silicon pillar meta-atoms on a silicon substrate with a period ($p$) of 58 μm. The height ($h$) of the meta-atom is fixed as 60 μm, and the width ($w$) and length ($l$) are independently tunable from 12 to 50 μm and 14 to 50 μm, respectively. This geometric tuning controls the local complex-amplitude response of a single meta-atom. When the wavefront passes through the metasurface, the near-field electric field distribution $E_0$ is obtained, which further propagates to a predefined target distance to reconstruct the target volume. As shown in Figure 1c, the holographic volume can range from a single-plane pattern to a 3D model.

The framework of LM-PINN is illustrated in Figure 1b, which consists of an input module, a network module, an output module, and a physical module. The input module provides two pathways to the network module: firstly, the target volume is directly divided into $N$ layers and fed into the network module (with a dimension of 64 × 64 × $N$); secondly, distance encoding: The electric field distribution $E_0$ at the metasurface plane is computed using existing 2D/3D holographic algorithms. The real and imaginary parts of $E_0$ are extracted and fed into the network model (with a dimension of 64 × 64 × 2). The network module is a Y-shaped convolutional neural network (termed Y-net), whose detailed structure is provided in Note S2. The Y-net outputs the $w$ map and $l$ map of the metasurface. A differentiable physical module, which integrates a local polynomial fitting process to map the metasurface geometry to the electric field distribution $E_{0,\mathrm{rec}}$ and the angular spectrum method (ASM) for wave propagation, connects the output of Y-net to the final reconstructed volume. Crucially, the loss function, which quantifies the difference between the input target volume and final reconstructed volume, is back-propagated to update the Y-net's weights. Compared with the supervised training of traditional DNNs, self-supervision only requires target image datasets to complete network training, without relying on paired input-label datasets (i.e., target and geometry datasets). After training, LM-PINN enables an end-to-end, target-to-geometry inference without pre-prepared geometry training data.

*2.1.1. Local polynomial fitting process*

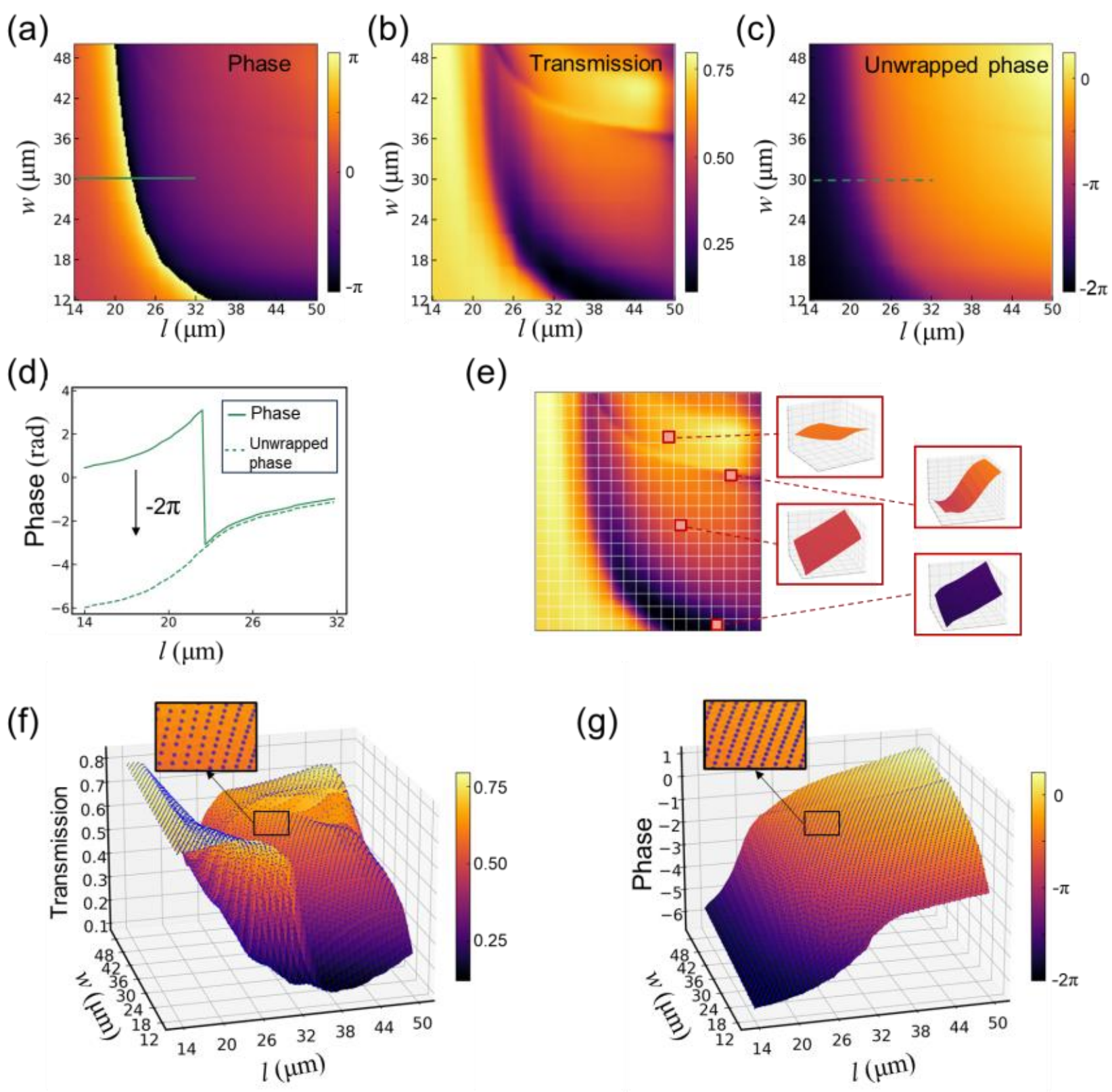

**Figure 2.** Local polynomial fitting. (a-b) FDTD simulation results of phase shift and transmission amplitude of a silicon meta-atom versus its width ($w$) and length ($l$). (c)

Unwrapped continuous phase distribution for subsequent local polynomial fitting. (d) Phase unwrapping process to eliminate $2\pi$ discontinuities. The solid and dashed green lines correspond to the cross-sectional profiles indicated by the green line counterpart in (a) and (c), respectively. (e) Schematic of the local polynomial fitting strategy. (f-g) Results of the local polynomial fitting for transmission and phase, respectively. The discrete blue points represent FDTD data, while the continuous surfaces represent the fitted results.

The local polynomial fitting process (**Figure 2**) serves as a highly efficient and differentiable surrogate model. It accurately maps the geometry ($l$ and $w$) of each meta-atom to its complex-amplitude E-field response (phase and transmission) at the near field, effectively replacing computationally intensive full-wave simulations during training. The model is constructed from exhaustive FDTD simulations of a single meta-atom. In the FDTD simulation, a THz plane wave with a wavelength of 116 μm is normally incident on the silicon substrate, passes through the meta-atom, and reaches the detector. Figure 2a and 2b present the phase and transmission response, respectively, obtained by parameter sweeps of the $w$ (12-50 μm) and $l$ (14-50 μm) with a step of 0.2 μm. The phase response in Figure 2a exhibits inherent 2π discontinuities that complicate polynomial fitting. To address this, a phase-unwrapping process was applied, whereby a $2\pi$ shift was subtracted from data points satisfying the condition of $w < 36$ μm and phase > 0, as shown in Figure 2d. This operation converted the discontinuous data (Figure 2a) into a smooth and continuous surface (Figure 2c), which is beneficial for polynomial fitting.

Instead of global polynomial fitting across the entire parameter space, which would be inaccurate for the complicated value distributions, a local polynomial fitting strategy was adopted. As illustrated in Figure 2e, the parameter space is divided into small, localized regions with an array of 10 × 10 pixels. Within each region, a 5th-order polynomial of the form

$$Y = a_0 + a_1 l + a_2 w + a_3 l^2 + a_4 wl + a_5 w^2 + \cdots + a_{20} w^5 \quad (1)$$

is then independently fitted to the FDTD data, where $Y$ represents the transmission or phase distribution within the region, and $a_0$, $a_1$, $a_2$, ..., $a_{20}$ are the corresponding local fitting coefficients. This method transforms the fitting of the complicated global parameter space into a set of simpler and smoother local approximations (Figure 2e), thereby significantly improving fitting accuracy. The results of the local polynomial fitting of the transmission and phase are presented in Figure 2f and Figure 2g, respectively. The discrete blue dots in the magnified insets represent the FDTD data, while the continuous surfaces show the values predicted by the local polynomial model. The close agreement between the two is quantified by remarkably low Normalized Mean Squared Error (NMSE) of $2.77\times10^{-6}$ for transmission and $2.17\times10^{-6}$ for phase. In contrast, the global fitting approach yields significantly higher errors of $1.53\times10^{-2}$ and $1.34\times10^{-2}$, respectively. Ultimately, this highly accurate and fully differentiable surrogate model is crucial for the stable and effective training of the LM-PINN.

*2.1.2. Angular spectrum method*

The Angular Spectrum Method (ASM) is employed due to its numerical rigor and differentiability, enabling accurate wave propagation modeling.[40] As illustrated in Figure 1b, the ASM calculates the complex-amplitude E-field distribution from the metasurface plane (z = 0) to the reconstructed image planes (z = $d_1$, $d_2$, ..., $d_N$) and propagation is implemented sequentially. Starting from z = 0, the E-field is propagated to the first plane (z = $d_1$). The resulting E-field at z = $d_1$ then serves as the source for propagation to the next plane (z = $d_2$). This process continues until the E-field at all planes has been computed. The detailed mathematical formulation is described in the Methods section.

*2.1.3. Loss function*

The loss function is designed to quantify the discrepancy between the reconstructed multi-plane images and the input target images, thereby driving the self-supervised learning process. By minimizing this loss through backpropagation, the LM-PINN is iteratively guided to discover the optimal metasurface geometry that generates the desired multi-plane holograms. The total loss $L_{total}$ is computed as the mean of the individual loss values evaluated at each target plane (z = $d_1$, $d_2$, ..., $d_N$):

$$L_{total} = \frac{1}{N}\sum_{i=1}^{N} L^{(i)} \tag{2}$$

where $L^{(i)}$ denotes the composite loss calculated for the *i*-th plane. At each plane, a composite loss function $L^{(i)}$ is formulated to incorporate constraints on both image shape and optical efficiency, ensuring high-quality holographic imaging while maximizing light utilization (see Methods section).

*2.1.4 Distance encoding*

The role of the distance encoding is to enable a trained LM-PINN model to implement various holographic volumes at different target distances. We denote the LM-PINN integrated with distance encoding as Dist-LM-PINN. During training, the target holographic volume is set as a single image, i.e., $N$=1. Each training data sample consists of one target image and a random diffraction distance $d$ (3 mm < $d$ < 20 mm). The angular spectrum method is used to propagate the target image to the metasurface plane with a distance of -*d*. The real and imaginary parts of the electric field distribution $E_0$ are then extracted and fed into the network module. During inverse design, the electric field distributions of various 2D/3D holographic volumes at the metasurface can be obtained through traditional holographic algorithms, which include target distance information. Inputting these electric distributions into the trained network module yields the geometric structure of the metasurface. Thus, a single Dist-LM-PINN model trained under a single-plane holography scenario can achieve end-to-end design of metasurfaces for different diffraction distances and different holographic volumes (Section 2.5).

**2.2. Single-plane holography**

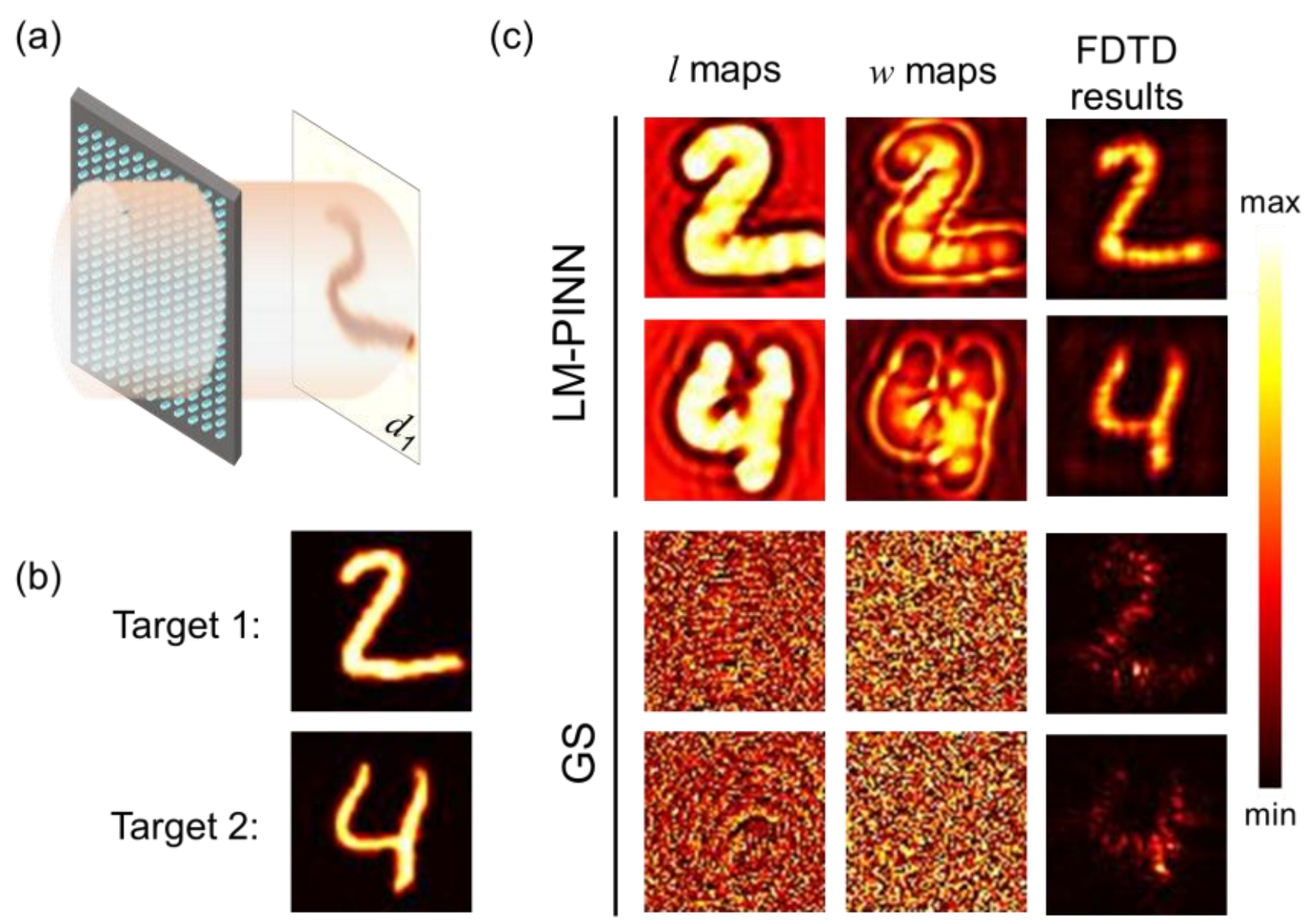

**Figure 3.** Simulation results of single-plane holography. (a) Scenario of metasurface-based single-plane holography. (b) Target images. (c) Length (*l*) map, width (*w*) map, and FDTD simulation results of LM-PINN and GS algorithms.

**Figure 3** presents the simulation results of single-plane holography (i.e., $N = 1$ in the $d_1$, $d_2$, ..., $d_N$ configuration illustrated in Figure 1). A schematic of the metasurface-based holography for this scenario is depicted in Figure 3a with the imaging plane located at $d_1 = 3$ mm. The MNIST datasets of handwritten numbers were chosen as the target images for training and testing (see Methods section). The numerals “2” and “4” serving as the two testing targets are shown in Figure 3b. The targets were input to the trained LM-PINN, which outputs the length (*l*) and width (*w*) maps (Figure 3c). For comparison, the GS algorithm was also employed to design the metasurfaces (3000 iterations). It is worth noting that while the GS algorithm outputs a phase map, which was subsequently converted into *l* and *w* maps via the mapping relationship, the LM-PINN is capable of implementing the end-to-end inverse design of the metasurface. The $l$ and $w$ maps were utilized in the full-wave FDTD simulation to verify the imaging performance of the metasurface (see Methods section). As illustrated in Figure 3c, the images reconstructed by LM-PINN exhibit superior reconstruction quality, showing more uniform intensity distributions inside the target regions. In contrast, the results of the GS algorithm suffer from serious discontinuities and blurred shape boundaries. Quantitative evaluations further confirm the high performance of LM-PINN, using the target images as ground truth to calculate the metrics of PSNR, SSIM, and NPCC (Table 1).

**Table 1.** Quantitative evaluation metrics of LM-PINN and GS algorithms in single-plane holography.

| Evaluation metrics | Algorithms | | | |
|---|---|---|---|---|
| | LM-PINN | | GS | |
| | Target 1 | Target 2 | Target 1 | Target 2 |
| NPCC | 0.971 | 0.964 | 0.778 | 0.804 |
| SSIM | 0.528 | 0.631 | 0.281 | 0.321 |
| PSNR | 17.9 dB | 19.6 dB | 10.8 dB | 13.2 dB |

The GS algorithm only processes phase information, with the assumption that the amplitude information is fixed. An additional step is required to map the phase information onto the metasurface structure ($l$ and $w$ maps). However, this mapping is not reversible, so the metasurface structures designed via the GS algorithm yield non-uniform amplitude distributions. And such neglect of amplitude information directly leads to a degradation in imaging quality. In comparison, the LM-PINN algorithm can directly output the $l$ and $w$ maps of the metasurface, enabling direct manipulation of the complex amplitude of light waves. Moreover, due to the strategy of random phase initialization, the $l$ and $w$ maps derived from the GS algorithm show no direct correlation with the target image (Figure 3c). In contrast, the metasurface structures designed via the LM-PINN algorithm exhibit a certain degree of correlation with the target images (Figure 3c). These are the reasons why the performance of the LM-PINN algorithm is superior to that of the GS algorithm.

**Figure 4** presents the experimental setup and results for holographic imaging using metasurfaces designed by LM-PINN. As illustrated in Figure 4a, a collimated beam from a THz quantum cascade laser (QCL) lasing at the frequency of 2.58 THz illuminates the metasurface, which is mounted on a translation stage. The intensity distribution of the reconstructed image is then recorded by a THz array detector (Swiss Terahertz, Rigi camera). The photograph and scanning electron microscopy (SEM) images of the fabricated metasurface are displayed in Figure 4c. The experimental imaging results exhibit a uniform structure and complete contour, accurately reproducing the preset shapes "2" and "4" with high fidelity (Figure 4d). The quantitative analysis for Target 1 and Target 2 is as follows: NPCC of 0.943 and 0.953, SSIM of 0.446 and 0.544, and PSNR of 15.1 dB and 17.8 dB, respectively. The imaging quality of the experimental results of LM-PINN is slightly degraded compared with that from the FDTD simulation results. This may be attributed to the relatively high background noise in the experimental results (Figure 4d), which is caused by multiple factors such as experimental thermal noise, optical path errors, and limited sensitivity of the THz array detector.

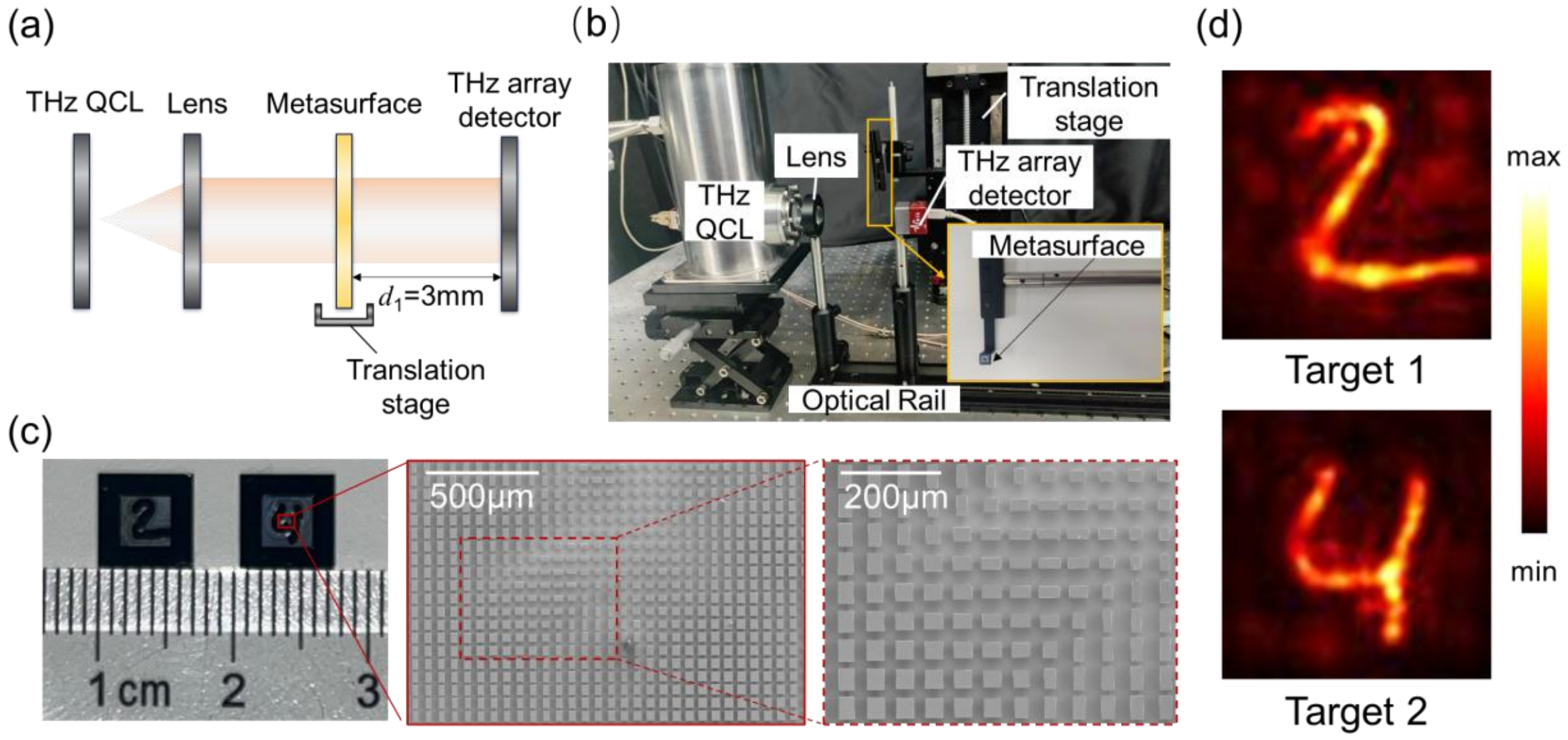

**Figure 4.** Experimental results of single-plane validation. (a) Schematic of the experimental setup. (b) Photograph of the experimental setup. (c) Photograph and SEM images of the fabricated metasurfaces. (d) Measured intensity distributions of the target images.

### 2.3. Multi-plane validation with an identical target image

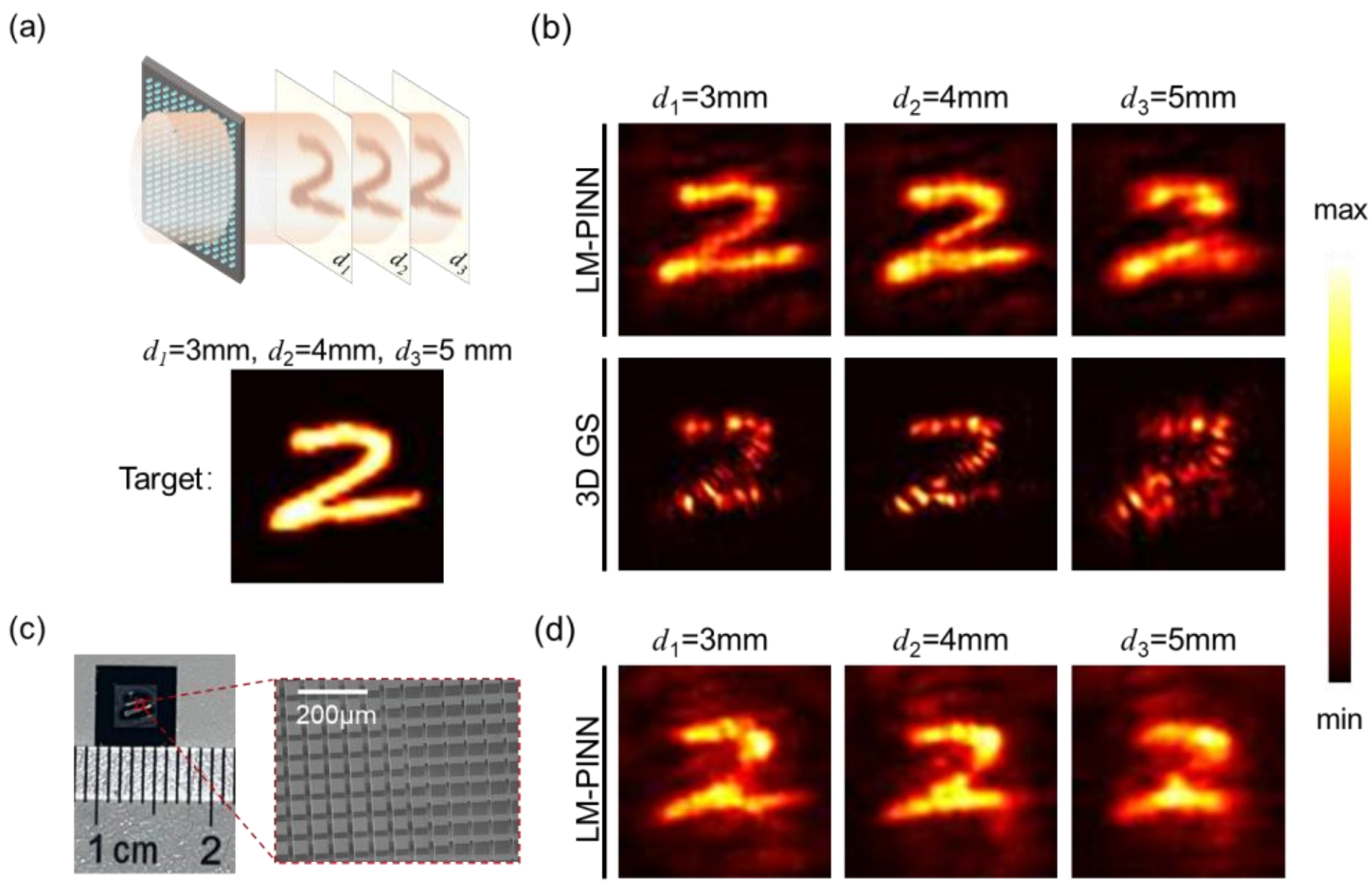

**Figure 5.** Multi-plane 3D validation with an identical target image. (a) Scenario of multi-plane meta-holography and the target image. (b) FDTD simulation results. (c) Photograph and SEM image of the fabricated metasurface. (d) Experimental results of the multi-plane THz holography.

**Figure 5**a illustrates the operating scenario of multi-plane metasurface-based holography, where the LM-PINN-designed metasurface is configured to reconstruct identical target images across the planes located at distinct axial distances ($d_1$ = 3 mm, $d_2$ = 4 mm, $d_3$ = 5 mm). It should be noted that the number and spacing of the imaging planes can be flexibly adjusted according to practical scenarios. Figure 5b shows the FDTD simulated results of the LM-PINN algorithm and 3D GS algorithm (3000 iterations) at corresponding axial distances.[21,41] The LM-PINN algorithm clearly reconstructs the outline of the number "2" on each imaging plane, and the light intensity distribution is in high agreement with the target. In contrast, the results of the GS exhibit poor pattern continuity, with the imaging quality deteriorating drastically, especially at $d_3$ = 5 mm. Quantitatively, the average metrics across the three planes of LM-PINN are: 0.957 (NPCC), 0.601 (SSIM), and 17.6 dB (PSNR). On the other hand, the average metrics of GS decrease to 0.861 (NPCC), 0.402 (SSIM), and 12.6 dB (PSNR).

To further validate the practical applicability of the LM-PINN algorithm, the experiments were subsequently conducted. The fabricated metasurface designed by LM-PINN is shown in Figure 5c, with a side-view SEM image clearly revealing the fabricated meta-atoms with high quality on the metasurface. The experimental setup employed herein is consistent with that depicted in Figure 4a-b. Specifically, the metasurface was translated forward and backward along the light propagation direction to acquire the multi-plane imaging data. As shown in Figure 5d, the experimental results exhibit not only a high consistency with the simulated counterparts but also a strong correlation between the measured patterns and the intended targets, thus verifying the feasibility of multi-plane holography using the LM-PINN-designed metasurface. And for the experiments, the average metrics across the three planes are: 0.924 (NPCC), 0.446 (SSIM), and 15.8 dB (PSNR), respectively. These results collectively demonstrate that the LM-PINN algorithm can be applied to achieve end-to-end design of high-quality 3D holographic metasurfaces.

### 2.4. Dual-plane validation with different target images

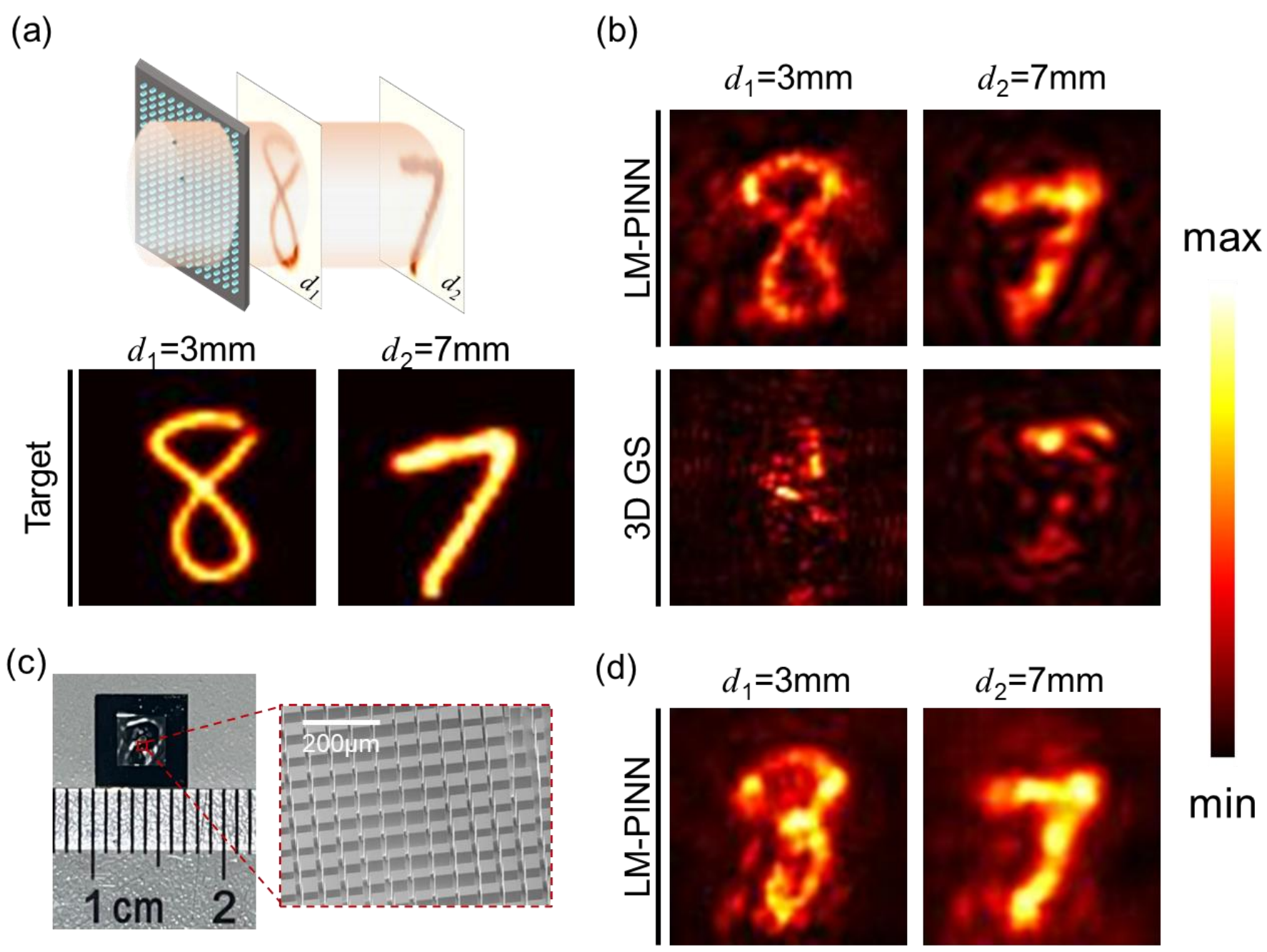

**Figure 6.** Dual-plane validation with different target images. (a) Scenario of dual-plane meta-holography with different target images. (b) FDTD simulation results. (c) Photograph and SEM image of the fabricated metasurface. (d) Experimental results of dual-plane THz holography with different target images.

**Figure 6** demonstrates the dual-plane 3D imaging capability of the LM-PINN-designed metasurface with different target images. Figure 6a illustrates the operating scenario: the metasurface is engineered to generate the digit "8" at $d_1 = 3$ mm and "7" at $d_2 = 7$ mm. Figure 6b presents the FDTD simulation results of the LM-PINN algorithm and 3D GS algorithm (3000 iterations). Compared with multi-plane 3D holography with identical patterns, the imaging performance of LM-PINN in this scenario degrades to a certain extent, which is mainly reflected in the increased background noise and reduced uniformity of light intensity in the target pattern regions (Figure 6b). Nevertheless, while the GS algorithm completely fails in this complex scenario, the LM-PINN algorithm can still reconstruct the expected pattern shape.

The experiments were then conducted, and Figure 6c shows the fabricated metasurface designed by the LM-PINN algorithm. The corresponding experimental results and the target patterns on both planes are successfully reconstructed, as shown in Figure 6d. Table 2 lists the quantitative evaluation results of the LM-PINN algorithm for both simulation and experiment. Compared with simple physical scenarios (e.g., single-plane holography and multi-plane holography with identical targets), the quantitative evaluation metrics of this scheme exhibit a certain decline, which is inherently determined by diffraction-based field shaping under multiple axial constraints. Nevertheless, some metrics even outperform those of FDTD simulation results of the GS algorithm in simple physical scenarios. This comparison demonstrates the powerful capability of the LM-PINN algorithm in designing metasurfaces for manipulating both simple and complex light fields.

**Table 2.** Quantitative evaluation metrics of LM-PINN in dual-plane holography.

| Evaluation metrics | FDTD simulation | | Experiment | |
|---|---|---|---|---|
| | $d_1$ = 3 mm | $d_2$ = 7 mm | $d_1$ = 3 mm | $d_2$ = 7 mm |
| NPCC | 0.903 | 0.923 | 0.839 | 0.908 |
| SSIM | 0.460 | 0.465 | 0.308 | 0.343 |
| PSNR | 17.5 dB | 17.3 dB | 15.1 dB | 15.4 dB |

**2.5. Universality of Single Trained Dist-LM-PINN for Diverse Scenarios**

The distance encoding is introduced (Figure 1b), allowing a single trained Dist-LM-PINN model to adapt to multiple physical scenarios. This model was trained for 10 epochs using the MNIST dataset with an image size of 64 × 64, and the training process is described in Section 2.1.4. It should be emphasized that the Dist-LM-PINN model with distance encoding was trained under the single-plane holography scenario. The well-trained Dist-LM-PINN model essentially establishes a mapping from the target electric field to the metasurface structure. Therefore, we only need to calculate the target electric field distribution ($E_0$) at the metasurface plane for the desired hologram using 2D/3D holographic algorithms,[42,43] and feed it as input to the Dist-LM-PINN, enabling the rapid design of the metasurfaces.

*2.5.1 Inverse design with tunable target distances*

To verify the universality of Dist-LM-PINN under different physical scenarios, holographic metasurfaces with various target distances were designed using Dist-LM-PINN, as illustrated in **Figure 7**a. For each target distance, the target images were adopted to calculate the electric field distribution ($E_0$) at the metasurface plane via the ASM. The electric field $E_0$ was then fed into the trained Dist-LM-PINN, which output the geometric parameters of the metasurface. FDTD simulation results of the designed metasurfaces are presented in Figure 7b. It can be seen that the metasurfaces designed by Dist-LM-PINN can realize holographic projection of target patterns at different target distances. In addition, the second and third rows of Figure 7b correspond to EMNIST and FashionMNIST datasets, respectively, demonstrating that the model trained on MNIST data possesses favorable generalization capability. The trained Dist-LM-PINN can also be applied to the design of patterns with larger sizes (512 × 512) and higher complexity (Note S3).

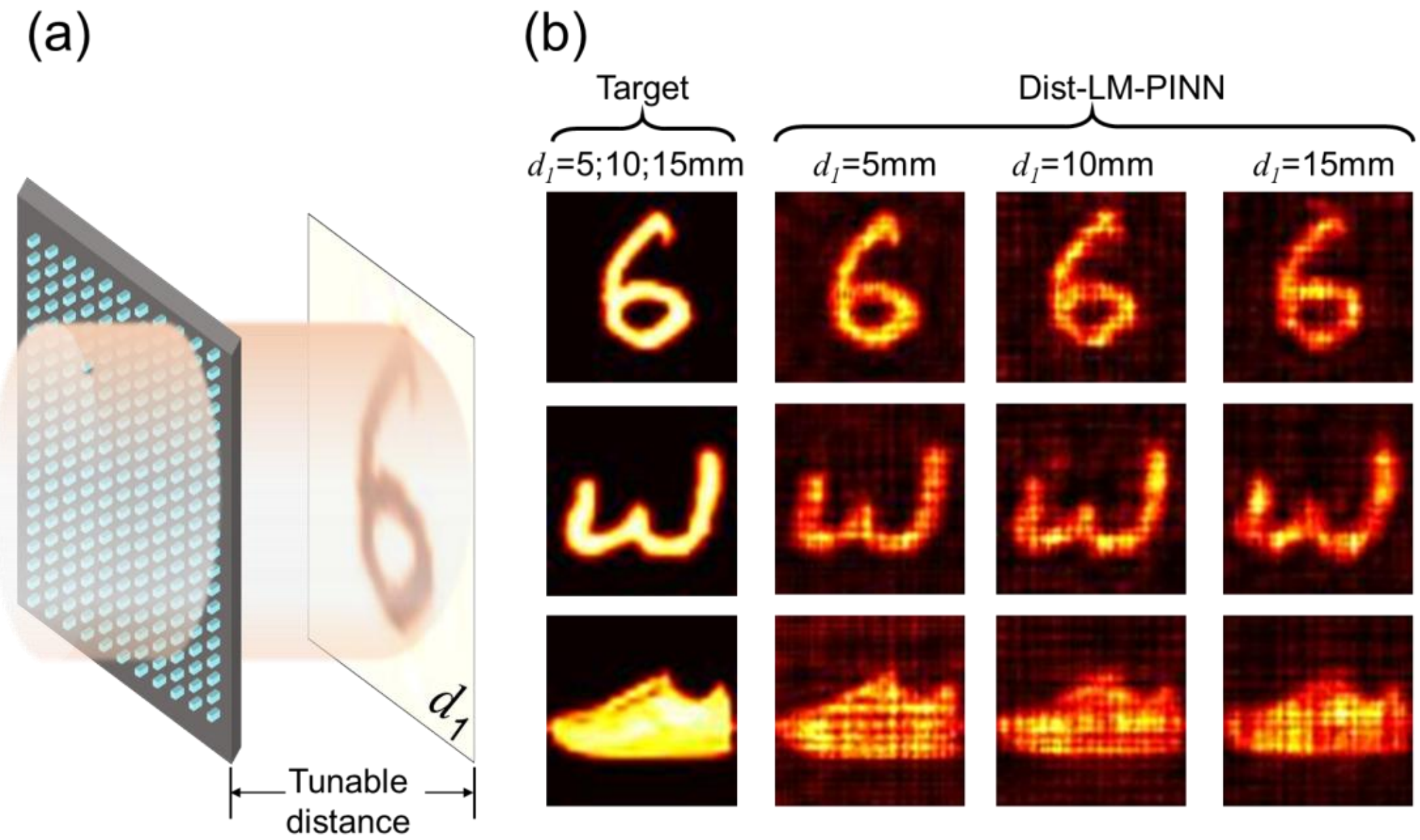

**Figure 7.** Single plane holography with a tunable distance. (a) Scenario of single plane holography with a tunable distance. (b) FDTD simulation results.

*2.5.2 Design of multi-focal metalens*

As shown in **Figure 8**a, the trained Dist-LM-PINN can also be adopted to design multi-focal metasurfaces. The target electric field distribution $E_0$ at the metasurface plane is calculated according to Reference [42] and input to the Dist-LM-PINN. To enhance the wavefront modulation capability of the metasurface, we set the size of the target electric field to 128 × 128. As illustrated in Figure 8b, two focal points are configured, with focal lengths $f_1$ = 4 mm and $f_2$ = 8 mm, respectively. FDTD simulation results of Dist-LM-PINN are displayed in Figure 8c. According to the x-z cross-section profiles and x-y cross-section profiles, the light intensity distribution of the metalens designed by Dist-LM-PINN is in excellent agreement with the target light intensity distribution.

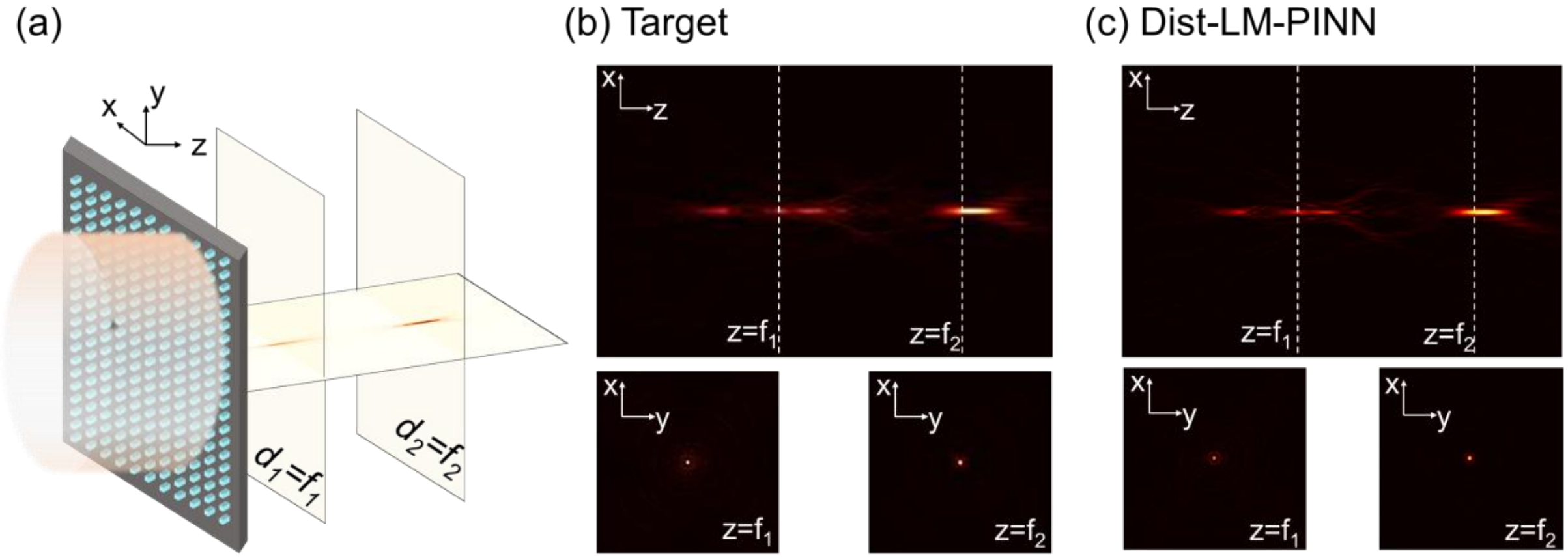

**Figure 8**. Design of multi-focal metalens. (a) Scenario of multi-focal metalens. (b) The target light intensity distribution. (c) FDTD simulation results of Dist-LM-PINN.

*2.5.3 Design of 3D holographic metasurface*

A single trained Dist-LM-PINN model can realize the multi-plane validation with identical target images described in Section 2.3 (Note S4). It enables flexible adjustment of the number of planes and inter-plane distances without retraining the model. In addition, this Dist-LM-PINN model can also handle the dual-plane validation scenario with distinct target images described in Section 2.4 (Note S4). To further verify the capability of Dist-LM-PINN for complex 3D holographic projection, an aircraft model is adopted as the target 3D hologram, as illustrated in **Figure 9**a. First, the 3D model is sliced into 15 cross-sections along the z-axis, with an inter-section spacing of 1 mm. The distance from the first cross-section to the metasurface is set to $d_1$ = 3 mm. The target electric field distribution $E_0$ at the metasurface is calculated via a 3D holographic algorithm,[43] which is then fed into Dist-LM-PINN to accomplish metasurface design. FDTD simulation results of the 15 cross-sections are presented in Figure 9b, and the corresponding 3D reconstruction result is shown in Figure 9c. These results demonstrate that Dist-LM-PINN is capable of designing sophisticated 3D holograms. Nevertheless, relatively strong background noise can be observed in Figure 9c, which is mainly attributed to the coupling effects between meta-atoms during FDTD simulations. Meanwhile, since the target electric field $E_0$ is derived from calculations of multiple layers, inter-layer crosstalk inevitably occurs, originating from the preprocessing 3D holographic algorithm.[43]

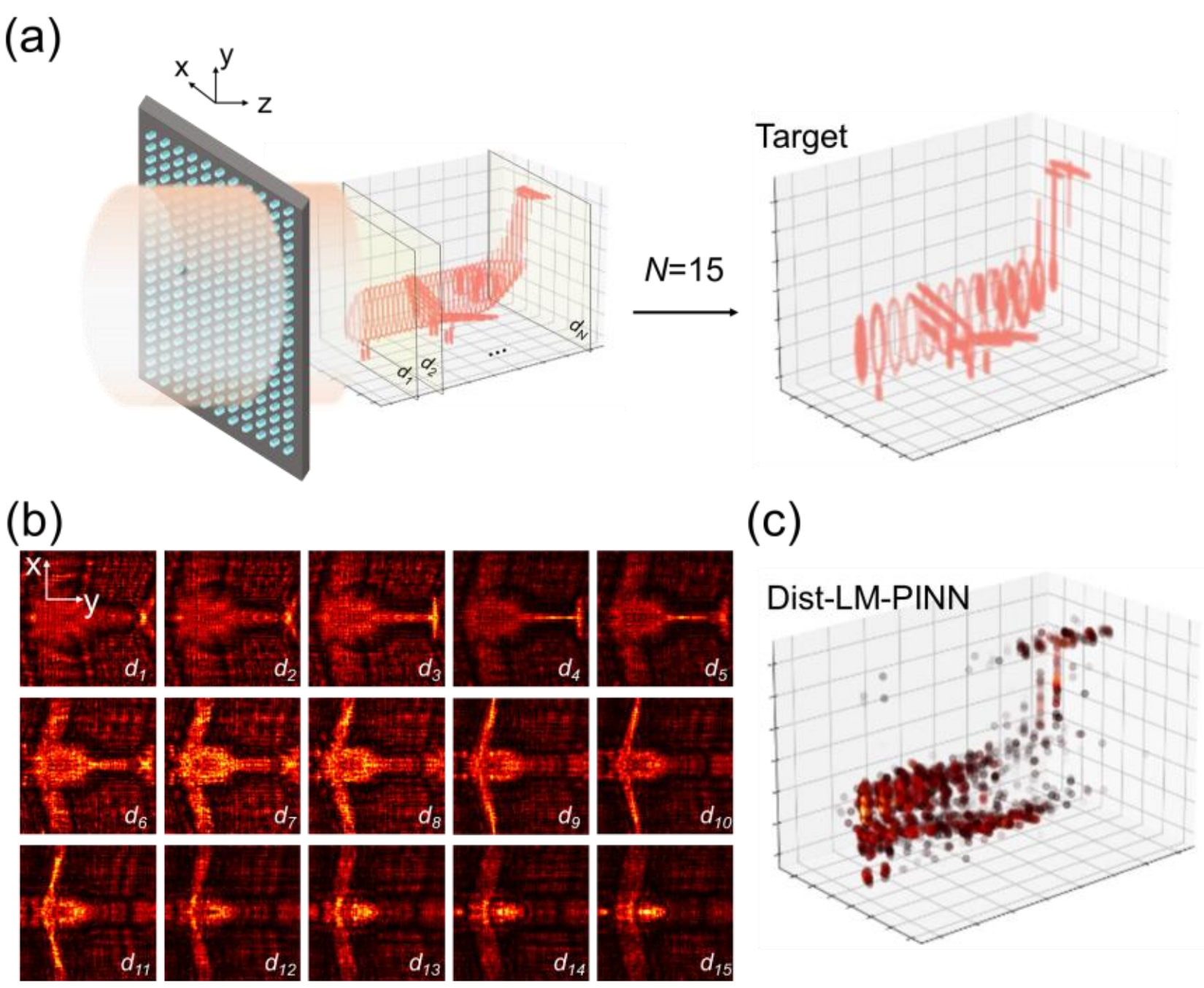

**Figure 9.** Design of 3D holographic metasurface. (a) Scenario of a metasurface-based 3D hologram. (b) FDTD simulation results in the x-y plane. (c) 3D reconstruction result of Dist-LM-PINN.

### 2.6. Computational efficiency

Table 3 presents the computation time of the LM-PINN and GS algorithms across the three aforementioned physical scenarios of metasurface-based THz holography. PINNs require considerable computational resources for network training, and the training time increases as the metasurface area grows. Fortunately, the LM-PINN enables the direct deployment of a network pre-trained on 64 × 64 structures for the design of large-area metasurfaces (Note S3). It can be observed that the LM-PINN algorithm achieves higher computational efficiency (i.e., less inference time) than the GS algorithm under all scenarios during metasurface design. With the introduction of distance encoding, satisfactory imaging quality (Section 2.5) can be achieved with only 10 training epochs (~0.4 hours). Benefiting from the strong generalization and versatility brought by distance encoding, the one-time training cost of the LM-PINN can be fully offset by its repeated reuse in various metasurface design tasks. For single-plane electric field inputs of size 512 × 512, the inference takes only 0.5 seconds with GPU acceleration. It should be noted that distance encoding itself also incurs a small computational overhead, typically on the order of tens of milliseconds (~0.02 seconds). Therefore, the LM-PINN lays a solid foundation for large-area and real-time 3D displays.

**Table 3.** The computational time of the LM-PINN and GS algorithms.

| | | Size of the metasurface | Without distance encoding | | | | | | With distance encoding |
|---|---|---|---|---|---|---|---|---|---|
| | | | Single-plane target | | Dual-plane (2 planes) with different targets | | Multi-plane (3 planes) with identical target | | Single-plane electric field |
| | | | LM-PINN | GS | LM-PINN | 3D GS | LM-PINN | 3D GS | LM-PINN |
| Train | GPU | 64 × 64 | 3.66h (100 epochs) | \ | 4.22h (100 epochs) | \ | 4.23h (100 epochs) | \ | 0.40h (10 epochs) |
| Infer | CPU | 64 × 64 | 0.38s | 1.31s | 0.38s | 2.61s | 0.38s | 3.99s | 0.41s |
| | | 512 × 512 | 1.12s | 14.90s | 1.18s | 30.02s | 1.19s | 43.89s | 1.17s |
| | GPU | 64 × 64 | 0.51s | 1.04s | 0.51s | 2.08s | 0.52s | 3.08s | 0.58s |
| | | 512 × 512 | 0.63s | 3.36s | 0.60s | 6.81s | 0.66s | 10.33s | 0.50s |

## 3. Conclusion

In conclusion, we demonstrated a physics-informed neural network (LM-PINN) framework for the rapid inverse design of complex-amplitude THz holographic metasurfaces. The LM-PINN architecture integrates four synergistic modules: an input module, a network module, an output module, and a physical module. The input module adopts a dual-pathway strategy, evolving from direct volumetric partitioning to an enhanced distance-encoding mechanism. This architectural advancement leads to the development of Dist-LM-PINN, which effectively transcends the "one-model-one-scenario" limitation prevalent in existing frameworks. Within this framework, the network module utilizes a Y-net architecture to directly generate the geometric parameter maps (length $l$ and width $w$) of the metasurface meta-atoms, facilitating a high-speed, end-to-end design process. The physical module incorporates a local polynomial fitting process and the multi-plane angular spectrum method. The former accurately maps geometric parameters to complex-amplitude responses with high accuracy, while the latter serves as a differentiable propagator to calculate plane-by-plane light distributions. The performance of the proposed LM-PINN has been validated through both numerical simulations and experimental characterization using a THz quantum cascade laser system. By leveraging distance encoding, a single pre-trained Dist-LM-PINN enables zero-shot generalization across diverse physical configurations, including varying target distances, adjustable plane numbers, and distinct 2D or 3D targets, without retraining. Compared with the traditional iterative GS algorithm, the LM-PINN achieves faster design speed (~1 second with CPU and ~0.5 seconds with GPU for a 512 × 512 metasurface) and generates higher-quality holographic images. This work will benefit the development of high-quality, large-scale 3D holographic display technology in the THz band. Looking forward, this strategy can be extended to incorporate additional degrees of freedom, such as operating wavelength, polarization state, etc. When combined with dynamic metasurfaces, which are tunable by these inputs, it holds great potential for realizing the rapid inverse design of photonic devices, thereby achieving real-time 3D holographic display.

## 4. Methods

### 4.1. Training details

The proposed LM-PINN framework was implemented using Python 3.11.9 and PyTorch 2.1.0. All models were trained on a workstation equipped with an Intel Xeon W-2235 CPU, 64 GB of RAM, and an NVIDIA GeForce RTX 3090 GPU. The training and testing datasets were derived from the MNIST handwritten digit dataset. A total of 10,000 training images and 100 test images were used. All images were resized to 64 × 64 pixels to match the metasurface dimensions, and the pixel values were normalized to a range from 0.01 to 1. For different experimental validations, the corresponding input training datasets (without labels) were constructed accordingly: (1) for the single-plane validation, the training set is directly constructed by randomly arranging 10,000 preprocessed MNIST images, and the test set is constructed in the same manner with 100 images; (2) for the multi-plane validation with identical target image, the same MNIST image was assigned to all propagation planes; (3) For the dual-plane validation with distinct target images, 10,000 preprocessed MNIST images were first duplicated to form two datasets. The two datasets were then randomly shuffled separately and used as the target patterns for plane 1 and plane 2, respectively, to construct the training set. The test set was prepared in the same way. The LM-PINN was trained for 100 epochs using the Adam optimizer with a learning rate of 0.001 and a batch size of 128.

### 4.2. Loss function

At each plane, the composite loss function $L^{(i)}$ is defined by:

$$L^{(i)} = L_{shape}^{(i)} + \alpha L_{light}^{(i)} \tag{3}$$

where $L_{shape}^{(i)}$ is the shape loss term, $L_{light}^{(i)}$ is the light efficiency loss term, and $\alpha$ is a weighting factor that balances the two constraints. The weighting factor $\alpha$ is set to 0.01. The shape loss term $L_{shape}^{(i)}$ combines Mean Squared Error (MSE) and Normalized Pearson Correlation Coefficient (NPCC) to enforce both pixel-wise accuracy and structural similarity between the reconstructed and target images:

$$L_{shape}^{(i)} = L_{MSE}^{(i)} + L_{NPCC}^{(i)} \tag{4}$$

$$L_{MSE}^{(i)} = \frac{1}{M}\sum\left(I_{tar}^{(i)} - I_{rec}^{(i)}\right)^2 \tag{5}$$

$$L_{NPCC}^{(i)} = 1 - \frac{\sum\left(I_{tar}^{(i)} - \bar{I}_{tar}^{(i)}\right)\cdot\left(I_{rec}^{(i)} - \bar{I}_{rec}^{(i)}\right)}{\sqrt{\sum\left(\left(I_{tar}^{(i)} - \bar{I}_{tar}^{(i)}\right)^2\cdot\left(I_{rec}^{(i)} - \bar{I}_{rec}^{(i)}\right)^2\right)}} \tag{6}$$

Here, $I_{tar}^{(i)}$ is the target image intensity, $I_{rec}^{(i)}$ is the reconstructed image intensity obtained via ASM, and $M$ is the total number of pixels. $\bar{I}_{tar}^{(i)}$ and $\bar{I}_{rec}^{(i)}$ represent the mean intensities of the target and reconstructed images, respectively.

The light efficiency loss $L_{light}$ is defined as:

$$L_{light}^{(i)} = 1 - \frac{\sum ln\left(I_{tar}^{(i)}\cdot I_{rec}^{(i)} + 1\right)}{\sum ln\left(I_{incident}^{(i)}\right)} \tag{7}$$

where $I_{incident}^{(i)}$ represents the incident light intensity (assumed to be a plane wave with unit intensity). The logarithmic operation reduces the internal contrast variations within the target region to ensure continuity.

### 4.3. Angular spectrum method (ASM)

Mathematically, the ASM can be formulated as

$$E_{z=d+\Delta d} = \mathcal{F}^{-1}\left[H \cdot \mathcal{F}(E_{z=d})\right] \tag{8}$$

Here, $E_{z=d}$ and $E_{z=d+\Delta d}$ represent the scalar components of the E-field at the spatial coordinates ($x$, $y$, $d$) and ($x$, $y$, $d$+$\Delta d$) in 3D space, respectively. $\mathcal{F}$ signifies the 2D Fourier transformation, which maps the spatial coordinates ($x$, $y$) to the spatial frequency coordinates ($f_x$, $f_y$). The transfer function $H(f_x, f_y)$ that describes the THz wave propagation from the plane z = $d$ to z = $d$ +$\Delta d$ is calculated as:

$$H\left(f_x, f_y\right)\Big|_{\Delta z=\Delta d} = exp\left(j\frac{2\pi}{\lambda}\Delta d\right)\cdot exp\left[-j\pi\lambda\Delta d(f_x^2 + f_y^2)\right] \tag{9}$$

In this expression, $\lambda$ denotes the wavelength and $\Delta d$ denotes the propagation distance between two planes along the optical axis (i.e., z-axis as shown in Figure 1). At each plane, the imaging pattern is the intensity distribution of the E-field:

$$I_z = |E_z|^2 \tag{10}$$

**4.4. Finite-difference time domain (FDTD) simulation**

Firstly, the FDTD method (Lumerical FDTD Solutions) was applied to calculate the phase shift and transmission data by sweeping geometric parameters of the meta-atom. In the simulation, a silicon pillar meta-atom was modeled on a silicon substrate. Periodic boundary conditions were applied in the x and y directions, while perfectly matched layers (PMLs) were configured at both the top and bottom boundaries. The height of the silicon pillar was fixed at 60 μm, with the width ($w$) ranging from 12 to 50 μm and the length ($l$) ranging from 14 to 50 μm at a step size of 0.2 μm. A plane wave source ($\lambda$ = 116 μm, $f$ = 2.58 THz, x-polarized) was positioned inside the silicon substrate and beneath the meta-atom, and a frequency-domain field and power monitor was placed above the meta-atom to detect the phase and transmitted power.

Secondly, the FDTD method was applied to calculate the imaging results of the metasurface designed by LM-PINN and GS algorithms. In the simulation, a 64 × 64 array of meta-atoms was arranged on a silicon substrate with a period of 58 μm ($\lambda/2$). The length ($l$) and width ($w$) parameters of each meta-atom were calculated by either the LM-PINN algorithm or the GS algorithm. All boundaries were configured as PMLs. A plane wave source ($\lambda$ = 116 μm, $f$ = 2.58 THz) was placed inside the silicon substrate beneath the metasurface, and a two-dimensional frequency-domain field and power monitor was deployed at the target imaging plane above the metasurface to record the electric field distribution. For 3D holographic imaging simulations, the monitor was placed at each target imaging plane. A monitor was also placed at the subwavelength plane behind the metasurface to measure its electric field distribution $E_{0,FDTD}$. As Section 2.6 involves FDTD simulations for long-distance, large-area metasurfaces, such simulations are extremely computationally demanding. Direct full-scale FDTD simulations of the entire physical scenario can take several days and may even lead to insufficient computational resources. Therefore, in Section 2.6, we first recorded the near-field distribution of the metasurface $E_{0,FDTD}$, and then propagated it to the target plane via the angular spectrum method to obtain the electric field distribution at the target position. The obtained electric field data at the target planes were subsequently used to calculate the light intensity distribution generated by the metasurface.

**4.5. Metasurface fabrication:**

The metasurface was fabricated on a 4-inch high-resistivity silicon wafer (double-side polished, resistance > 10000 Ω, thickness = 500 μm) via a series of microfabrication processes. Firstly, the silicon wafer was cleaned on a wet cleaning station. Subsequently, pattern transfer was achieved by a contact lithography system with ultraviolet exposure (SUSS-MA8). Deep silicon etching was then carried out by Deep Reactive Ion Etching (DRIE) technique using $SF_6$ and $C_4F_8$ as process gases, targeting an etch depth of 60 μm (error < 10%). After etching, the photoresist was removed. All metasurface patterns were fabricated uniformly on the silicon wafer. Finally, each metasurface was diced from the silicon wafer by a dicing saw.

**4.6. Experimental setup:**

The metasurface was characterized with a THz quantum cascade laser (QCL) operating at a wavelength of 116 μm (2.58 THz) with an average output power of ~1.4 mW. The emitted THz wave was collimated into a parallel beam using a TPX lens with a focal length of 25 mm, and was subsequently incident on the metasurface. A THz array detector (Swiss Terahertz, Rigi camera) was employed to capture the intensity distribution. The array detector features a pixel size of 25 μm and an effective detection area of 3 mm × 4 mm. To fully characterize the multi-plane reconstruction, the camera was sequentially positioned at the respective reconstructed planes ($z = d_1, d_2, ..., d_N$). Since the incident collimated beam was smaller than the effective modulation area of the metasurface, the sample was mounted on an XYZ-translation stage. By spatially scanning the metasurface relative to the stationary beam, the THz signals from different regions were captured and subsequently stitched together to form a complete intensity image at the imaging plane.

**Acknowledgements**

The work is supported by the National Natural Science Foundation of China under Grants 62375170 and 62535019, Shanghai Jiao Tong University under Grant No. YG2024QNA51, the Science and Technology Commission of Shanghai Municipality under Grant No. 20DZ2220400.

**Conflicts of Interest**

The authors declare no conflicts of interest.

**Data Availability Statement**

The data and codes that support the findings of this study are available upon reasonable request.

# Supporting Information

**Metasurface-based Terahertz Three-dimensional Holography Enabled by Physics-Informed Neural Network**

*Jingzhu Shao, Ping Tang, Borui Xu, Xiangyu Zhao, Yudong Tian, Yuqing Liu, and Chongzhao Wu**

J. Shao, P. Tang, B. Xu, X. Zhao, Y. Tian, Y. Liu, C. Wu
Center for Biophotonics
Institute of Medical Robotics
School of Biomedical Engineering
Shanghai Jiao Tong University,
Shanghai 200240, China
*Email: czwu@sjtu.edu.cn

**Note S1: Existing THz holography enabled by complex-amplitude metasurfaces.**

**Table S1.** Comparison of our method with existing Terahertz holography enabled by complex-amplitude metasurfaces.

| Ref. | Frequency | Design method | Speed | 3D display | Sample | PSNR | Efficiency[#] | Correlation Coefficient |
|---|---|---|---|---|---|---|---|---|
| Optics and Laser Technology (2024)[3] | 0.6~1.4 THz | Metasurface structures capable of independently modulating phase and amplitude | End-to-end | \ | English letters "B" "I" "T" | \ | \ | 0.85-0.88 |
| *ACS Photonics* (2018)[1] | 1 THz | | End-to-end | Yes | English letters "T" | \ | 16.3% | \ |
| *Advanced Materials Technologies* (2025)[2] | 0.009 THz | | End-to-end | Yes | English letters | 14.3 dB | \ | \ |
| *ACS Appl. Mater. Interfaces* (2022)[4] | 0.01 THz | Forward design with a DNN for mapping the electronic response of the meta-atom to its parameters. | End-to-end | \ | English letters "A" "I" | 14.88 ~ 16.51 dB | \ | \ |
| *Photonics Res.* (2025)[5] | 0.3~1.2 THz | Unsupervised physics-informed DNN with an inverse network and a pretrained forward network | iterative | Yes | English letters "S" | 11.91 ~ 12.39 dB | \ | \ |
| our method | 2.58 THz | Inverse design based on self-supervised physical-informed DNN | End-to-end | Yes | Arabic numerals "2" "4" "7" "8" | 15.1~17.8 dB | 30.6%~40.7 % | 0.92~0.95 |

[#] Efficiency is defined as the ratio of the terahertz energy within the target pattern region to the incident terahertz energy.

Although complex-amplitude metasurfaces often achieve superior holographic performance compared with phase-only metasurfaces, their design is generally more complicated. One common approach is to first calculate the complex-amplitude distribution of the metasurface from the target pattern via diffraction calculations, then map each pixel of the complex-amplitude distribution to an individual meta-atom. This typically requires metasurface structures capable of independently modulating phase and amplitude[1–3]. An alternative strategy is to train a neural network that realizes the mapping from electromagnetic responses to metasurface unit structures.[4] Similar to Ref. 5, our method incorporates a physical model into network optimization, enabling direct mapping from target patterns to metasurface structures. The difference is that Ref. 5 requires a pre-trained forward prediction network to map metasurface structures to their electromagnetic responses, and the metasurface structure is obtained through iterative optimization. In contrast, our method replaces the forward prediction network with local polynomial fitting, and the trained network enables end-to-end on-demand design. Table S1 summarizes the comparison between our method and these reported complex-amplitude holographic metasurfaces operating in the terahertz band. Since different evaluation metrics are adopted for implementation assessment, we additionally calculated the Efficiency and Correlation Coefficient.

**Note S2: Detailed architecture of the Y-net.**

As shown in Figure S1, the network module is a Y-shaped convolutional neural network (termed Y-net). The Y-net consists of one encoder pathway for processing input data, followed by two decoders that directly output the *w* map and *l* map of the metasurface, respectively.

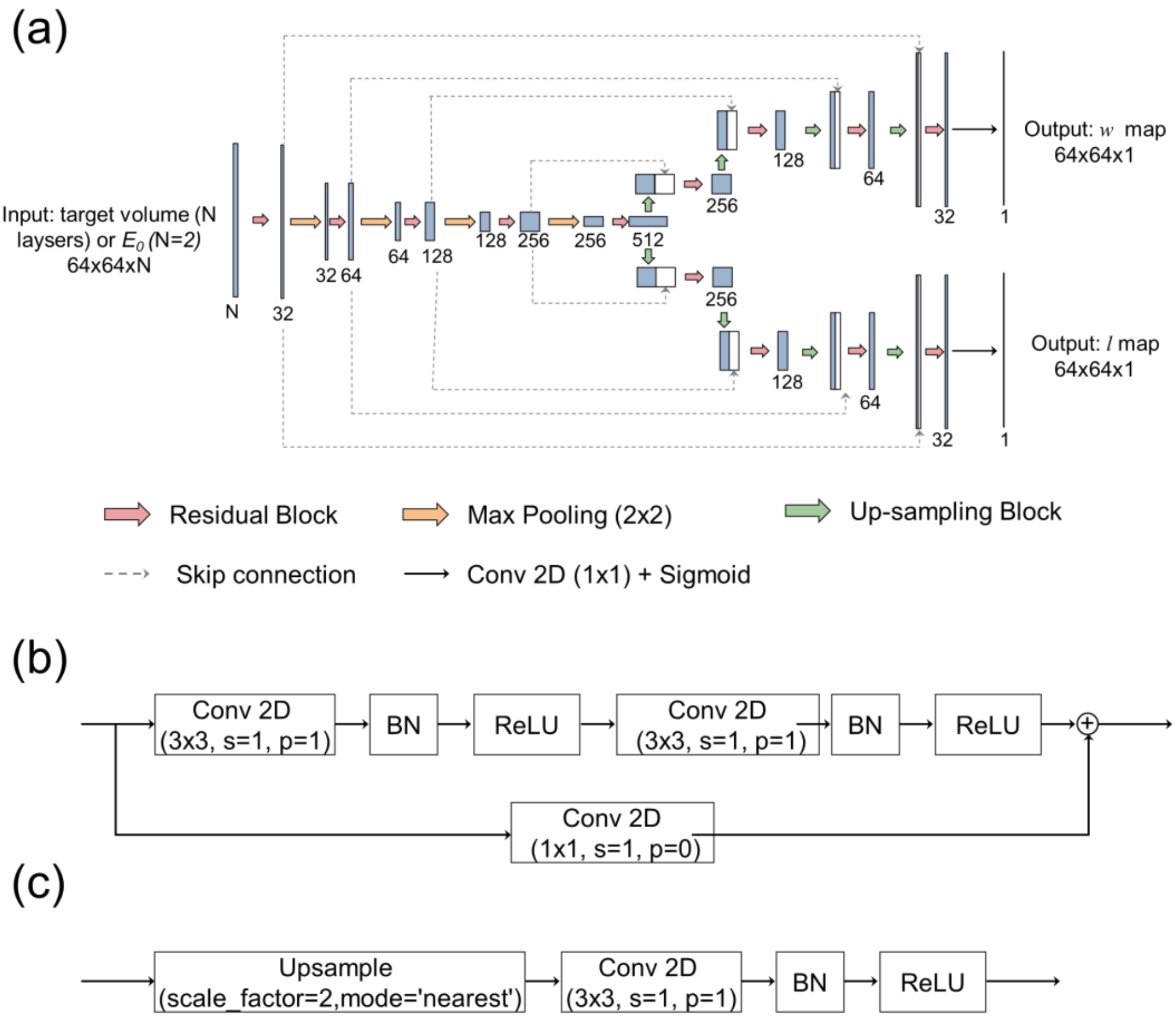

**Figure S1.** (a) Detailed structure of the Y-net. (b) Residual Block. (c) Up-sampling block.

**Note S3 Design of metasurface for large (512 × 512) and complex target image using Dis-LM-PINN.**

As shown in Figure S2a, the target holographic pattern is the emblem of Shanghai Jiao Tong University, with a resolution of 512 × 512 pixels and a target imaging distance of 25 mm. After calculating the target electronic field $E_0$ via the Angular Spectrum Method (ASM), the $E_0$ is fed into the Dist-LM-PINN model to obtain the corresponding *w* map and *l* map of the metasurface (Figure S2b and Figure S2c). Notably, the Dist-LM-PINN model is trained on the 64 × 64 MNIST dataset, and the target distance range during training is 3-20 mm. FDTD simulation of the 512 × 512 metasurface requires enormous computing resources (more than hundreds of GB of memory). Therefore, in Figure S2d, we directly present the reconstruction result of Dist-LM-PINN obtained by ASM propagation (i.e., the "reconstructed volume" shown in Figure 1b of the manuscript). It can be observed that the reconstructed result is in excellent agreement with the target pattern.

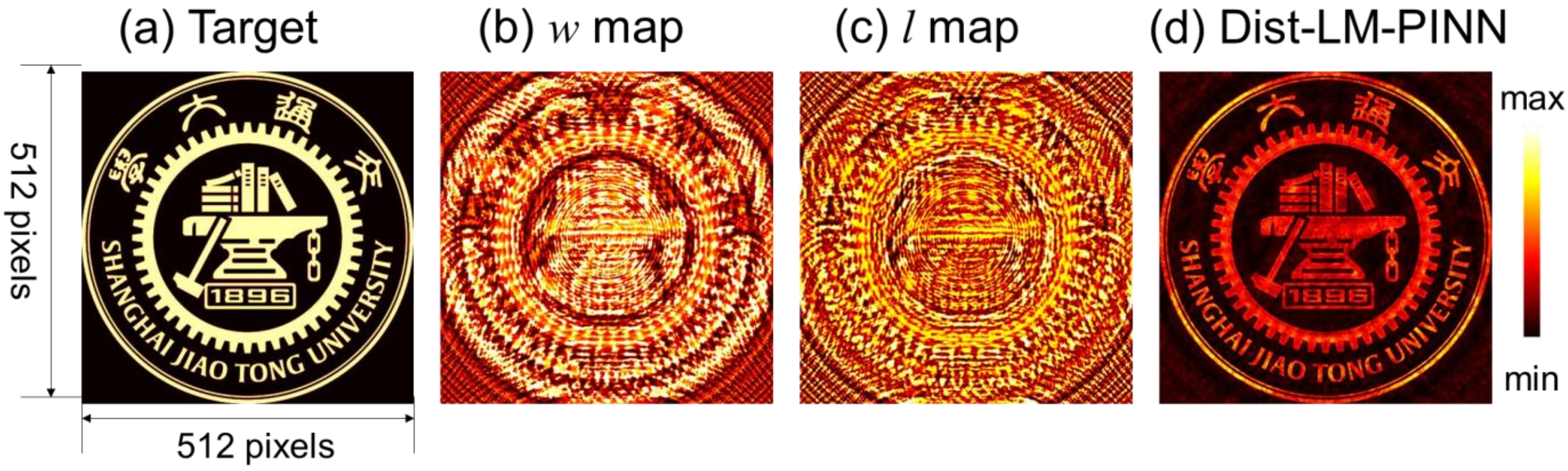

**Figure S2.** Design of metasurface for large (512 × 512) and complex target images based on Dis-LM-PINN. (a) Target image. (b) *w* map. (c) *l* map. (d) Reconstructed image of Dist-LM-PINN.

**Note S4 Design of multi-plane holographic metasurface using Dist-LM-PINN.**

A single well-trained Dist-LM-PINN model (integrated with distance encoding) is capable of performing the functions of multiple individually trained conventional LM-PINN models (without distance encoding). The input of the Dist-LM-PINN model is the electric field, which is obtained by averaging the electric field distributions directly back-propagated from each target plane to the metasurface. The electric field on each target plane is constructed by taking the square root of the target pattern intensity distribution as the electric field amplitude, with the phase uniformly set to zero. As illustrated in Figure S3, the single Dist-LM-PINN model can not only reproduce the results of LM-PINN (Section 2.3 in the manuscript), but also realize multi-plane 3D holography with arbitrary numbers of target planes and arbitrary inter-plane distances. As shown in Figure S4, Dist-LM-PINN can also implement the dual-plane validation scenario with different target images (Section 2.3 in the manuscript).

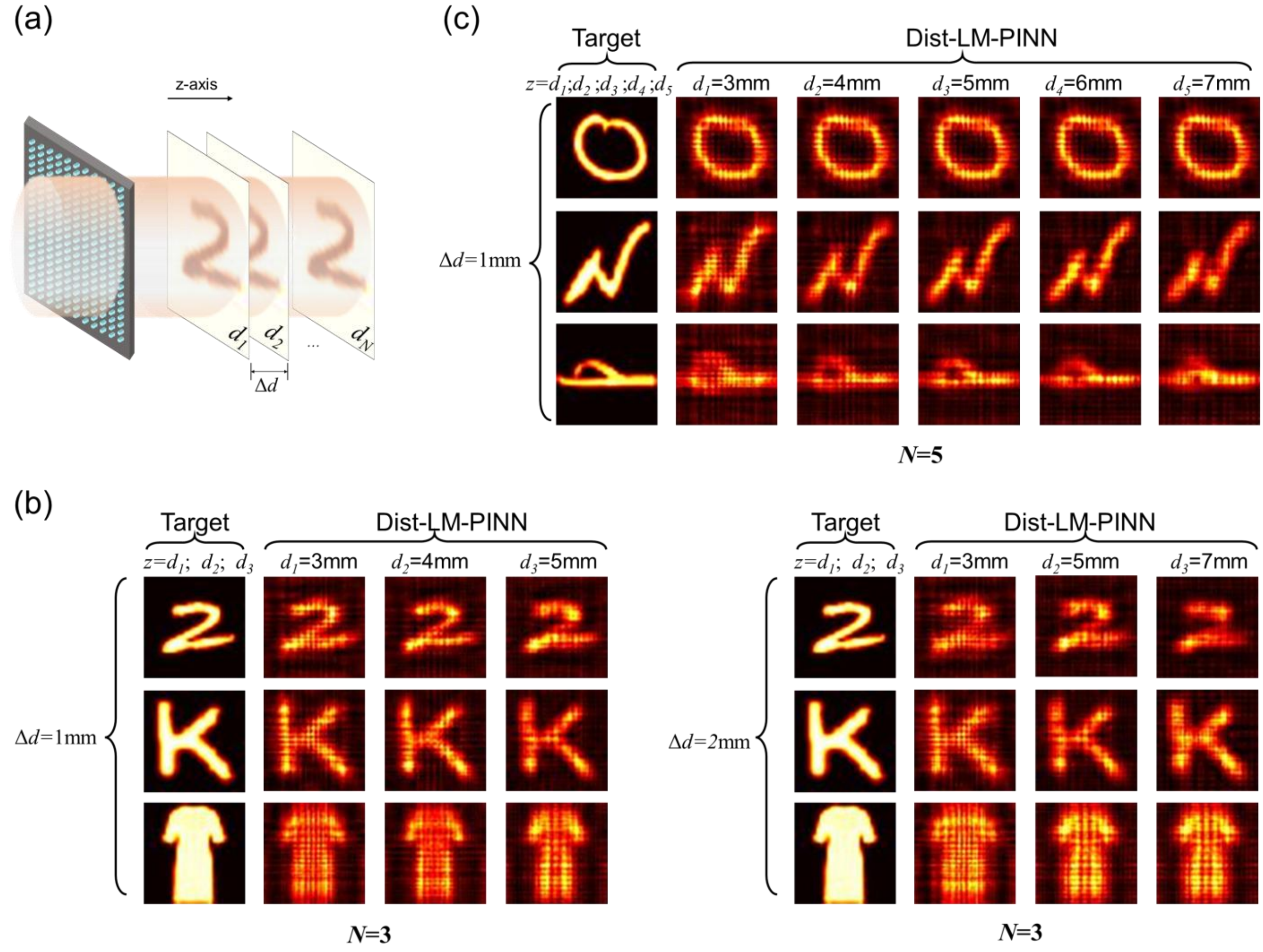

**Figure S3.** Multi-plane 3D validation with an identical target image. (a) Scenario of multi-plane meta-holography. (b) FDTD simulation results of three-plane holography. (c) FDTD simulation results of five-plane holography.

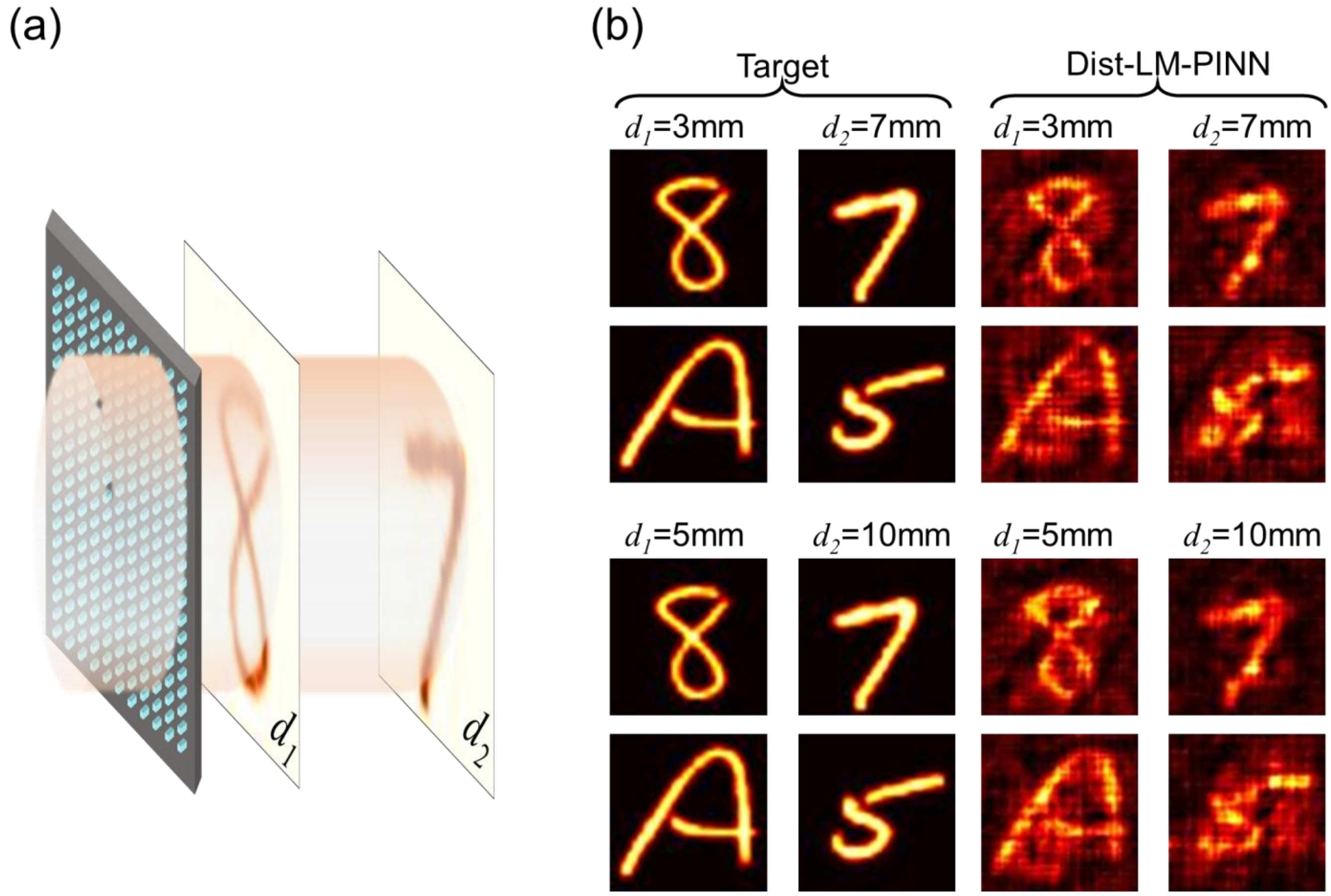

**Figure S4.** Dual-plane validation with different target images. (a) Scenario of dual-plane meta-holography with different target images. (b) FDTD simulation results.